\documentclass[nofootinbib, letterpaper, floatfix, preprintnumbers, twocolumn]{revtex4}  
\pdfoutput=1
\usepackage{amsmath,amssymb}
\usepackage{cancel}
\usepackage{mathrsfs}
\usepackage{graphicx}
\usepackage{hepunits}
\usepackage{wasysym}
\usepackage[utf8x]{inputenc}
\begin{document}
\title{Lepton Flavor and Non-Universality
from Minimal Composite Higgs Setups
}
\author{Adri\'an Carmona} 
\affiliation{Theoretical Physics Department,
CERN, 1211 Geneva 23, Switzerland}

\author{Florian Goertz}
\affiliation{Theoretical Physics Department,
CERN, 1211 Geneva 23, Switzerland}
\date{\today}

\preprint{CERN-PH-TH-2015-251}

\begin{abstract}
	We present a new class of models
of lepton flavor in the composite Higgs framework.
Following the concept of minimality,
they lead to a rich phenomenology in good agreement 
with the current experimental picture.
Because of a unification of the right-handed leptons,
our scenario is very predictive and can naturally lead to a violation
of lepton-flavor universality in neutral current interactions.
We will show that, in particular, the anomaly in 
$R_K= {\cal B}(B \to K \mu^+ \mu^-)/{\cal B}(B \to K e^+ e^-)$, 
found by LHCb, can be addressed,
while other constraints from quark- and lepton-flavor
physics are met.
In fact, the minimal structure of the setup allows for
the implementation of a very powerful flavor protection, which avoids the appearance of new sources of flavor-changing neutral 
currents to very good approximation.
Finally, the new lepton sector provides a parametrically
enhanced correction to the Higgs mass, such that the need
for ultra-light top partners is weakened considerably, linking the mass
of the latter with the size of the neutrino masses.

\end{abstract}
\maketitle

\section{Introduction}
Composite Higgs models (CHMs), where the Higgs boson is a bound state of some new strong interaction \cite{Kaplan:1983fs, Georgi:1984af},  offer a compelling framework for addressing the gauge hierarchy problem, since  the Higgs mass is protected by its finite size. In addition, if it is assumed to be the Nambu-Goldstone boson of some global symmetry of the strong sector \cite{Dimopoulos:1981xc, Contino:2003ve, Agashe:2004rs}, only broken by the weak couplings 
to the SM-like spin-1 and spin-1/2 elementary degrees of freedom (dof), the Higgs boson is naturally much lighter than the scale where the new strong interaction starts to be resolved. This allows in particular for a robust effective low-energy description without specifying its constituents.  If one further assumes linear mixings of the SM-like fields (besides the Higgs) with the composite operators,  the light mass eigenstates will be mixtures of elementary and composite dof,  naturally explaining the observed flavor structure. This concept is known as \emph{partial compositeness}. In this Letter, we present a very minimal
implementation of the lepton sector in CHMs, realizing neutrino masses
via a type-III seesaw mechanism. This allows to unify the right-handed (RH) lepton sector by embedding both the RH charged leptons as well as the RH neutrinos in a 
{\it single} representation of the global $SO(5)$ of the minimal composite Higgs model \cite{Agashe:2004rs} (MCHM), reducing the dof as well as the number of 
parameters with respect to standard realizations of the fermion sector. 

Interestingly, linked to this unification, our setup predicts a
violation of lepton-flavor 
universality (LFU) in neutral current interactions, while LFU is basically respected
in charged currents. In fact this is notable in the light of the recent LHCb measurement
\cite{Aaij:2014ora} of the ratio
\begin{equation}
\begin{split}
R_K&= \left. \frac{{\cal B}(B \to K \mu^+ \mu^-)}{{\cal B}(B \to K e^+ e^-)}\right|_{q^2 
\in [1,6]\,{\rm GeV}}^{\rm exp}\\[2mm]
&=0.745^{+0.090}_{-0.074}\pm 0.036\,,
\end{split}
\end{equation}
which shows a 2.6$\,\sigma$ discrepancy with respect to the Standard Model (SM) prediction of 
$\left|R_K^{\rm SM}-1\right|<1\%$ \cite{Hiller:2003js}. We will see that such a value can arise very 
naturally in our setup.

Finally, the non-trivial lepton sector has a significant impact on the composite-Higgs potential, allowing for a light 125\,GeV Higgs boson without the necessity of ultra-light top partners with masses below the Higgs-decay constant $f_\pi \lesssim 1\,$ TeV \cite{Carmona:2014iwa}. Here we will entertain for the first time a full flavor model, implementing in particular a flavor symmetry that will lead to a vanishing of potentially dangerous flavor-changing neutral currents (FCNCs).
We will particularly scrutinize the phenomenology in the (lepton) flavor sector, with a special emphasis on $R_K$.


\section{Setup, EWSB and Flavor Structure}
We consider the {\it minimal} custodial embedding of the SM lepton sector including three RH fermion triplets with zero hypercharge,  $\Sigma_{\ell R}$, with $\ell=e,\mu,\tau$. If these new dof have Majorana masses of order $\mathcal{O}(M_{\rm GUT})$,  the observed tiny neutrino masses can be explained with $\mathcal{O}(1)$ Yukawa couplings via the (type-III) \emph{seesaw} mechanism. In the framework of the MCHM, or its five dimensional (5D) holographic dual \cite{Maldacena:1997re,Gubser:1998bc,Witten:1998qj,ArkaniHamed:2000ds}, where the Higgs is identified with the pseudo Nambu-Goldstone bosons of $SO(5)/SO(4)$, this is realized 
by embedding all the RH leptons within each generation in the same symmetric representation ($\mathbf{14}$) of $SO(5)$, whereas every left-handed (LH) doublet is embedded in a fundamental representation ($\mathbf{5}$) of the same group.

This minimal realization of the lepton sector,  has some striking consequences on electroweak symmetry breaking (EWSB) and the mass of the Higgs boson. First of all, by unifying RH neutrinos and charged leptons within the same $\mathbf{14}$ of $SO(5)$, it links together several \emph{a priori} unrelated aspects of the theory, like the overall size of the neutrino masses with the breaking of the EW symmetry and the Higgs mass and thus finally with the masses of top partners.  This can be easily understood within the dual 5D  theory; since the Majorana masses can only arise from the ultra-violet (UV) brane, where $SO(4)$ is not respected, in order to warp down  the otherwise $\mathcal{O}(M_{\rm Planck})$ effective Majorana masses, the corresponding RH zero-modes have to be localized closer to the infra-red (IR) brane, leading then to RH charged leptons with a sizable degree of compositeness. (Note that custodial symmetry will still guarantee the agreement of the $Z \ell_R \bar{\ell}_R$ couplings with EW precision measurements.) Moreover, since the quantum numbers of the symmetric representation allow for the lepton contribution to the Higgs mass to appear at leading order in the degree of compositeness, contrary to the general case, large effects are expected even for moderately composite leptons. By partially canceling the top quark contribution to the Higgs mass in a sizable region of parameter space, this enhanced lepton contribution can then allow for a larger top-quark breaking of the Goldstone symmetry while still reproducing the light Higgs mass. This relaxes significantly the upper bound on the mass of the potentially light top partners (see Figure~\ref{fig:mhvsmtp} below), which would otherwise be required in order to reduce the Goldstone breaking for a fixed $m_t$. 
Finally, the possibility of having an additional source of sizable contributions to the Higgs mass allows for a viable EWSB in previously disregarded most minimal models where the contribution of the top-sector alone would not be sufficient (at least in scenarios that have a holographic dual). This is the case for a fully composite $t_R$, with the LH doublet being embedded in a $\mathbf{5}$, which we will employ in the following.

\begin{figure}[!b]
	\begin{center}
			\includegraphics[width=0.48\textwidth]{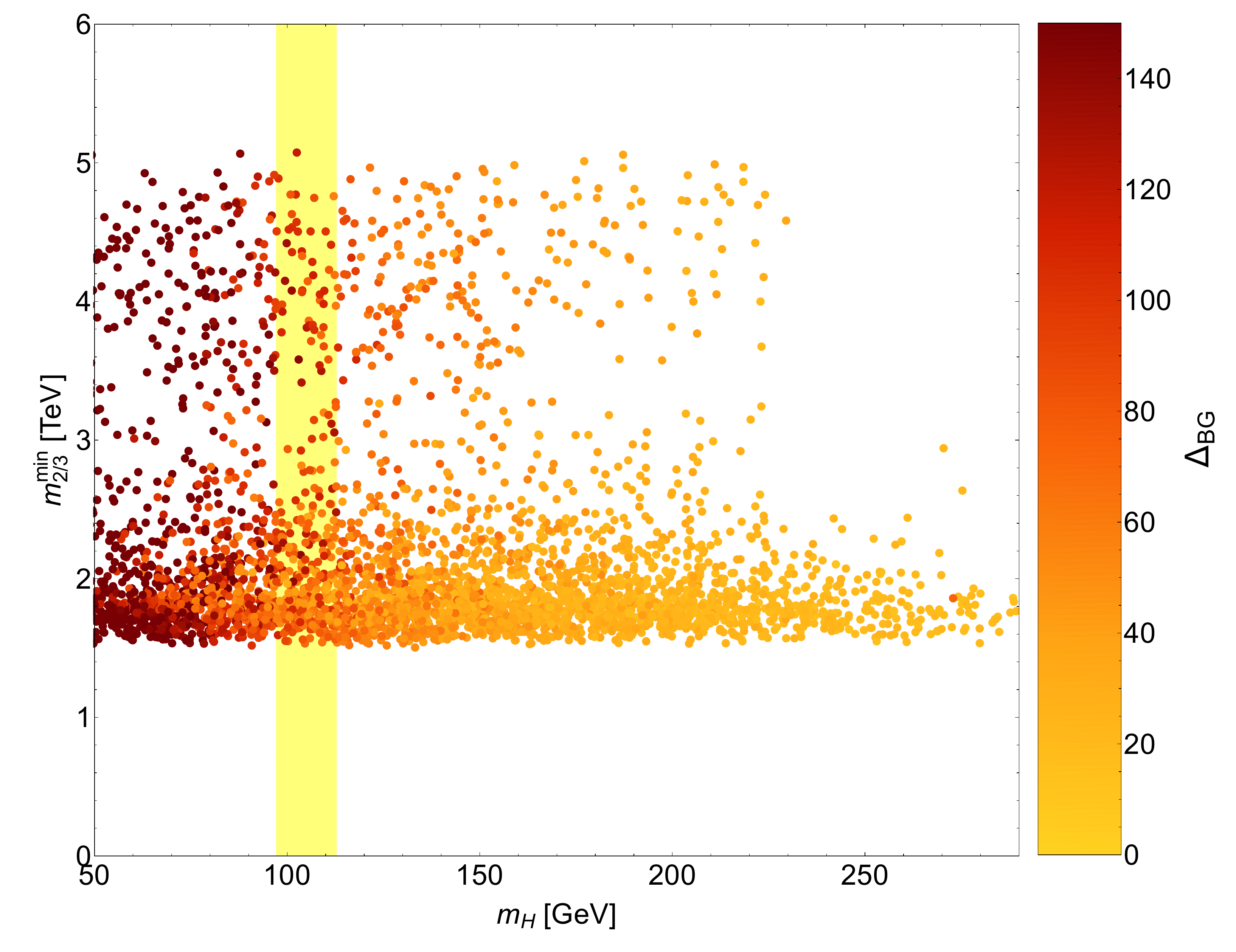}
		   \caption{Mass of the lightest top partner versus the Higgs mass as a function of the tuning $\Delta_{\rm BG}$, with lighter points corresponding to smaller values of $\Delta_{\rm BG}$, for $f_{\pi}=1\,$TeV. The yellow band corresponds to $m_H(f_{\pi})=105$~GeV~$(1\pm7.5\%)$.}
			\label{fig:mhvsmtp}
		\end{center}
\end{figure}

To be more specific, let us consider the following 5D holographic theory where the usual four-dimensional (4D) Minkowski space is extended with a slice of AdS$_5$, 
\begin{align}
	\mathrm{d}s^2&=a^2(z)\left[\eta_{\mu\nu}\mathrm{d}x^{\mu}\mathrm{d}x^{\nu}-\mathrm{d}z^2\right],\nonumber\\
	a(z)&=R/z,
\end{align}
where $z$ is the coordinate of the additional spatial dimension $R\le z\le R^{\prime}$ and $R\sim M_{\rm Pl}^{-1}$ 
($R^{\prime}\sim {\rm TeV}^{-1}$) is the position of the UV (IR) 4D brane. The global symmetry breaking $SO(5)/SO(4)$ of the 4D strongly coupled theory is realized in the 5D holographic picture by an IR brane respecting only an $SO(4)$ gauge symmetry, whereas the bulk along the extra dimension is gauge invariant under $SO(5)$. In order to reproduce the correct Weinberg angle, an extra $U(1)_X$ gauge symmetry is added to both the bulk and the IR brane. Finally, the gauging of $G_{\rm EW}=SU(2)\times U(1)_Y$ is realized in the 5D theory in a similar way, by keeping the UV brane just invariant under $G_{\rm EW}\subset SO(4) \times U(1)_X$. 

The lepton matter content consists of just the $SO(5)\times U(1)_X$ multiplets $\zeta_1^{\ell}\sim \mathbf{5}_{\mathbf{-1}},~ \zeta_2^{\ell}\sim\mathbf{14}_{-1}$ for $\ell=e,\mu,\tau$. To specify how the different zero-modes, identified with the SM particles, are embedded in these fields, we give the boundary conditions at the two 4D-branes
\begin{eqnarray}
\zeta_{1}^{\tau}&=&\tau^{\prime}_{1}[-,+]\oplus \left(\begin{array}{r}\nu_{1}^{\tau}[+,+]~  ~\tilde{\tau}_1[-,+]\\ \tau_{1}[+,+]~ \tilde{Y}_1^{\tau}[-,+]\end{array}\right),\nonumber\\
\zeta_{2}^{\tau}&=&\tau^{\prime}_{2}[-,-]\oplus\left(\begin{array}{r} \nu_{2}^{\tau}[+,-]~~  \tilde{\tau}_2[+,-]\\ \tau_{2}[+,-]~ \tilde{Y}_2^{\tau}[+,-]\end{array}\right) \\
						   &\oplus&\left(\begin{array}{r} \hat{\lambda}^{\tau}_2[-,-]~~\nu_{2}^{\tau\prime\prime}[+,-]~  ~~\tau_2^{\prime\prime\prime}[+,-]\\ 
	\hat{\nu}_2^{\tau}[-,-]~~~\tau_{2}^{\prime\prime}[+,-]~Y_2^{\tau\prime\prime\prime}[+,-]\\ \hat{\tau}_2[-,-]~Y_2^{\tau\prime\prime}[+,-]~\Theta_2^{\tau\prime\prime\prime}[+,-] \end{array}\right),\nonumber\qquad
\end{eqnarray}
where for simplicity we have focused on the third generation, with the other two being completely analogous. We have explicitly shown the decomposition under $SU(2)_L\times SU(2)_R$, where the bidoublet is represented by a $2\times 2$ matrix on which the $SU(2)_L$ rotation acts vertically and the $SU(2)_R$ one horizontally. The left and right columns correspond to fields with $T_R^3=\pm 1/2$, while the upper and lower rows have $T_L^3=\pm 1/2$ (and analogously for the bi-triplet). The hypercharge of a given fermion fulfills $Y=T_R^3+Q_X$, with $Q_X$  the $U(1)_X$ charge. The signs in square brackets denote the boundary conditions at the UV and IR branes. A Dirichlet boundary condition for the RH (LH) chirality is denoted by $[+]$ ($[-]$), with LH (RH) zero modes being present for fields with $[+,+]~ ([-,-])$ boundary conditions. The quark sector is given by the following four multiplets per generation: $\xi_1^{i}\sim \mathbf{5}_{2/3},\, \xi_2^i\sim \mathbf{1}_{2/3},\, \xi_3^{i}\sim \mathbf{5}_{-1/3},\,  \xi_4^i\sim \mathbf{1}_{-1/3}$,  $i=1,2,3$, 
or more explicitly, 
\begin{eqnarray}
\xi_{1}^i&\!=\!\left(\begin{array}{r}\tilde{\Lambda}^i[-,+]~  u^i_1[+,+]\\ \tilde{u}^i[-,+]~ d_1^i[+,+]\end{array}\right)\oplus u_1^{i\prime}[-,+],\quad \xi_{2}^i[-,-],\quad\\
\xi_{3}^i&\!=\!\left(\begin{array}{r}u^i_3[-,+]~\tilde{d}^i[-,+]  \\ d_3^i[-,+]~\tilde{\Xi}^i[-,+]\end{array}\right)\oplus d_3^{i\prime}[-,+],\quad \xi_{4}^i[-,-].\quad
\end{eqnarray}

This minimal model of composite leptons naturally allows for a very strong  flavor protection, requiring any lepton flavor violating (LFV) process to be mediated by extremely suppressed neutrino-mass insertions and leading, in particular, to the absence of dangerous FCNCs in the lepton sector
to excellent approximation. First of all, it is straightforward to see that in the conformal or decompactified limit, given by $R\to 0$ and $R^{\prime}\to \infty$, the lepton sector of the theory features an accidental $U(3)_1\times U(3)_2$ global symmetry corresponding to arbitrary rotations in the flavor space of the 5D multiplets $\xi_1^{\ell}$ and $\xi_2^{\ell}$. However, this global symmetry is broken in the compact theory by different values of the fermion bulk masses as well as by the presence of IR brane masses, required to generate a massive low-energy chiral spectrum. One possibility is to assume that these sources of flavor violation are controlled by the vacuum expectation value of some non-dynamical field $\mathcal{Y}$  \cite{Fitzpatrick:2007sa, Perez:2008ee}. Since we want these symmetries to be global symmetries of the strongly coupled 4D theory, we will consider the bulk of the extra dimension and the IR brane (its holographic dual) to be  $SU(3)_1\times SU(3)_2$ gauge invariant. The bulk fields will thus transform as $\zeta_1\sim (\mathbf{3},\mathbf{1}),~ \zeta_2\sim (\mathbf{1},\mathbf{3})$.
On the other hand, since the elementary sector represented by the UV brane does not respect this symmetry, one can have general Majorana masses
\begin{eqnarray}
\mathcal{L}_{\rm UV}\supset-\frac{1}{2}\left.a^4M_{\Sigma}^{\ell\ell^{\prime}}\mathrm{Tr}\left(\bar{\Sigma}_{\ell R}^c \Sigma_{\ell^{\prime} R}\right)\right|_{z=R}+\mathrm{h.c.},
	\label{eq:uvmass}
\end{eqnarray}
where
\begin{eqnarray}
\Sigma_{\ell}=\begin{pmatrix}\hat{\nu}_2^\ell/\sqrt{2}&\hat{\lambda}_2^\ell\\ \ell_2& -\hat{\nu}_2^\ell/\sqrt{2}\end{pmatrix}, \qquad \ell=e,\mu,\tau\,.
\end{eqnarray}
However, in the bulk and on the IR brane, the gauge flavor symmetry will be only broken by powers of the spurion $\mathcal{Y}\sim (\mathbf{3},\mathbf{\bar{3}})$.
Therefore, the corresponding bulk masses will be given by
\begin{eqnarray}
	c_1=\eta_1\mathbf{1}+\rho_1\mathcal{Y}\mathcal{Y}^{\dagger},\qquad c_2=\eta_2\mathbf{1}+\rho_2\mathcal{Y}^{\dagger}\mathcal{Y},
	\label{eq:bulkmasses}
\end{eqnarray}
while the IR brane masses will read
\begin{eqnarray}
	\left.a^4\left[ \omega_S\left(\bar{\zeta}_{1L}^{(\mathbf{1},\mathbf{1})}\mathcal{Y}\zeta_{2R}^{(\mathbf{1},\mathbf{1})}\right)+\omega_B(\bar{\zeta}_{1L}^{(\mathbf{2},\mathbf{2})}\mathcal{Y}\zeta_{2R}^{(\mathbf{2},\mathbf{2})})\right]\right|_{R^{\prime}}+\mathrm{h.c.}, 
	\label{eq:irmasses}
\end{eqnarray}
where the superscripts $(\mathbf{1},\mathbf{1})$ and $(\mathbf{2},\mathbf{2})$ denote the singlet and the bidoublet components, while $\eta_{1,2},\rho_{1,2}\in\mathbb{R}$ and $\omega_{S,B}\in\mathbb{C}$. The fact of having just two $SO(5)$ lepton multiplets and thus being able to use only one $SU(3)_1\times SU(3)_2$ spurion, $\mathcal{Y}$, allows us to diagonalize both (\ref{eq:bulkmasses}) and (\ref{eq:irmasses}) by performing the rotation $\zeta_1\to \mathcal{U}_1\zeta_1,~ \zeta_2\to \mathcal{U}_2\zeta_2$,
where $\mathcal{U}_1^{\dagger}\mathcal{Y}~\mathcal{U}_2=\mathrm{diag}(y_{ee},y_{\mu\mu},y_{\tau\tau})\equiv y_{\ell\ell}$.
Therefore, in this particular basis, the whole Lagrangian will be flavor diagonal with the exception of the Majorana mass in (\ref{eq:uvmass}), which becomes $\mathcal{U}_2^{T}M_{\Sigma} ~\mathcal{U}_2$.
Note that potential FCNCs induced by this structure will be suppressed by large Majorana masses.
We take random values of $|\omega_{S,B}|\le 1$, $0\le \eta_1\le 1$, $|\rho_1|\le 1$ as well as $|\Delta_{\rho}|=|\rho_1-\rho_2|\le 0.1$, obtaining $\eta_2$ and $0.1\le |y_{\ell \ell}|\le 0.7, ~\ell =e,\mu,\tau$, as a function of the charged lepton masses and the scale of the neutrino masses. 

For simplicity, we consider the more general case of arbitrary sources of flavor breaking in the quark sector, with their IR and UV brane masses reading
\begin{eqnarray}
- \left.a^4\left[M_{u}^{ij}\overline{\xi}^{i(\mathbf{1},\mathbf{1})}_{1L} \xi_{2R}^j+M_d^{ij}\overline{\xi}^{i(\mathbf{1},\mathbf{1})}_{3L} \xi_{4R}^j\right]\right|_{R^{\prime}}+\mathrm{h.c.},~
	\label{eq:irmassesq}
\end{eqnarray}
and $-\left.a^4\left[\Delta^{ij} \bar{q}_{1L}^i q_{3R}^j\right]\right|_{R}+\mathrm{h.c.}$, respectively,  
where $M_{u,d}^{ij}\in\mathbb{C}$. The  mixing terms  $\propto \Delta^{ij}\in \mathbb{C}$ are  needed in order to split the LH zero-mode between the up and the down sectors. For simplicity and to avoid ruining the hierarchical structure of the quark mass matrix \cite{Csaki:2008zd}, we assume that $\Delta=\delta \mathbf{1}$. We take random values $|M_{u,d}^{ij}|\le 0.7$, $|\delta|\le 0.2$ and select points which fit the observed quark masses and mixing angles at the 95\% C.L..

In order to get a quantitative idea of the impact of the lepton sector on the Higgs potential and the Higgs mass, we show in Figure~\ref{fig:mhvsmtp} the mass of the lightest top partner versus the Higgs mass evaluated at the composite scale $\mathcal{O}(f_{\pi})$, with the yellow band corresponding to the high-scale value of the actual Higgs mass $m_H(f_{\pi})=105$~GeV~$(1\pm7.5\%)$, after accounting for the uncertainties of the running in a conservative way. We also show the Barbieri-Giudice (BG) measure of the tuning   $\Delta_{\rm BG}=\mathrm{max}_{i,j}|\partial \log \mathcal{O}_i/\partial X_j|$, with $\mathcal{O}_i$ the observables considered and $X_j$ the different input parameters (see \cite{Carmona:2014iwa} for more details), via the color of each point in the $m_{H}-m_{2/3}^{\rm min}$ plane. $\Delta_{\rm BG}$ is defined to encode the amount of tuning among the physical parameters of the theory \cite{Barbieri:1987fn}, with values of $\Delta_{\rm BG}\sim 10,100$ corresponding to cancellations of one and two orders of magnitude, respectively.  Such tuning is expected to scale as $v^2/f_{\pi}^2$.  We can see from the figure that top-partner masses up to $5\,$TeV are allowed with a more than reasonable amount of tuning.

\section{Lepton Non-Universality and $R_K$}

Besides featuring a large impact on the Higgs potential  and the possibility of implementing a flavor protection leading to an almost FCNC-free setup with virtually no LFV, the unification of every generation of RH leptons in a single  $SO(5)$ multiplet leads to additional interesting consequences. The mixing of the elementary leptons with the composite sector is described in the dual 4D strongly coupled theory by the following Lagrangian 
	\begin{eqnarray}
		\mathcal{L}_{\rm mix}=\frac{\lambda_{L}^{\ell}}{\Lambda^{\gamma_L^{\ell}}}\bar{l}_{\ell L}\mathcal{O}_{\ell L}+\frac{\lambda_R^{\ell}}{\Lambda^{\gamma_R^{\ell}}}\bar{\Psi}_{\ell R}\mathcal{O}_{\ell R}+\mathrm{h.c.}
	\end{eqnarray}
where  $\Lambda$ is the UV cut-off scale, $\gamma_{L,R}^{\ell}=[\mathcal{O}_{\ell L,R}]-5/2$ are the different anomalous dimensions, $\lambda_{L,R}^{\ell}$ are order one dimensionless parameters and all RH leptons have been embedded in $\Psi_{\ell R}\sim \mathbf{14}$. Since we expect $\gamma_{R}^{\ell}<0$, a large contribution to the $\Psi_{\ell R}$ kinetic term will be generated at the scale $\mu=\mathcal{O}(\rm{TeV})$ where the conformal sector becomes strongly coupled, leading, after its normalization, to the following expressions for the physical masses, $\mathcal{M}_{e}\sim \delta_{\ell \ell^{\prime}} v\epsilon_{\ell L}$ and $\mathcal{M}_{\nu}\sim v^2 \epsilon_{\ell L}\epsilon_{\ell R}\left(M_{\Sigma}\right)^{-1}_{\ell \ell^{\prime}} \epsilon_{\ell^{\prime} L}\epsilon_{\ell^{\prime} R}$, where $\epsilon_{\ell L,R}\sim \lambda_{L,R}^{\ell}(\mu/\Lambda)^{\gamma_{L,R}^{\ell}}$ and $M_{\Sigma}$ is the elementary Majorana mass. It is then clear  that featuring at the same time hierarchical charged lepton masses and a non-hierarchical neutrino mass matrix \emph{requires} $\epsilon_{eL}\ll \epsilon_{\mu L}\ll \epsilon_{\tau L}\ll 1$ and $\epsilon_{\ell L} \epsilon_{\ell R}\sim \mathrm{constant}$, 
and thus $0 \ll \epsilon_{\tau R}\ll \epsilon_{\mu R}\ll \epsilon_{e R}$,
predicting, in particular,  violation of LFU. (See Refs.~\cite{Gripaios:2014tna, Niehoff:2015bfa} for different examples in the context of CHMs and Refs.~\cite{Crivellin:2015mga, Crivellin:2015lwa, Sierra:2015fma, Celis:2015ara, Falkowski:2015zwa} for other $Z^{\prime}$ models.)

For fermions with a sizable degree of compositeness, four-fermion operators mediated by vector resonances which may be relevant for electroweak precision data (EWPD) and flavor arise
$c\mathcal{O}\sim c\, (\bar{\psi}_2\gamma_{\mu} \psi_1)(\bar{\chi}_2\gamma^{\mu} \chi_1)$, where $c\sim \epsilon_{\psi_1}\epsilon_{\psi_2}\epsilon_{\chi_1}\epsilon_{\chi_2}/f_{\pi}^2$.

\begin{figure}[!h]
	\begin{center}
			\includegraphics[width=0.42\textwidth]{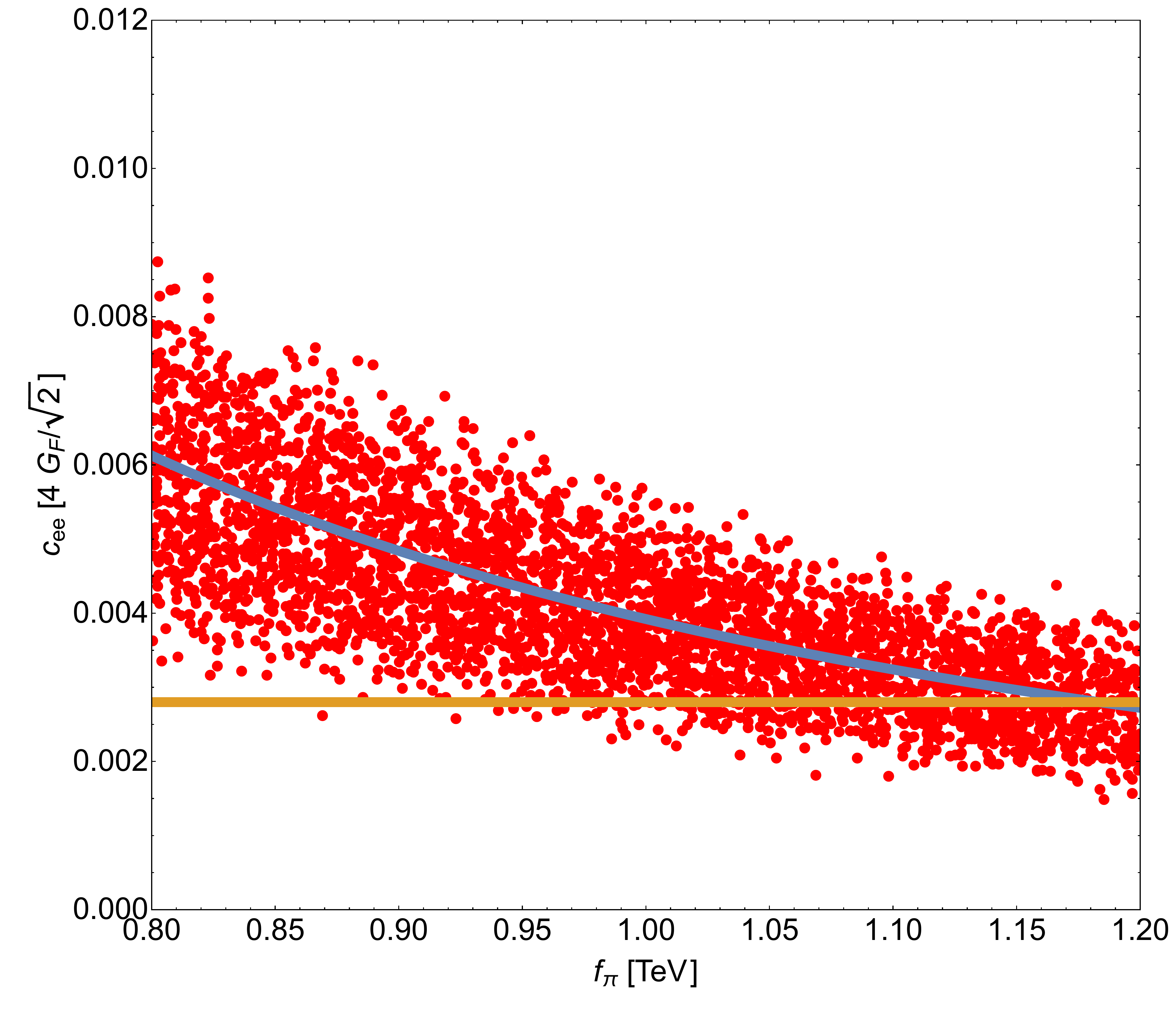}
			\caption{Value of $c_{ee}$ versus $f_{\pi}$. The blue curve shows the best fit to the data while the yellow line corresponds to the upper bound at 95\% C.L..}
			\label{fig:4fer}
		\end{center}
\end{figure}

According to the expected hierarchy in the RH mixings $\epsilon_{\ell R}$, the most important of these operators regarding EWPD will be $\mathcal{O}_{ee}=(e_R\gamma_{\mu} e_R)(e_R\gamma^{\mu} e_R)/2$, whose Wilson coefficient $c_{ee}$  is constrained to be  $c_{ee}\in 4 G_F/\sqrt{2}\cdot [-1.8,+2.8] \cdot 10^{-3}$ at 95\% C.L.  \cite{Raidal:2008jk}.
We show in Figure~\ref{fig:4fer} the value of $c_{ee}$ versus $f_{\pi}$, where the blue curve corresponds to the best fit to the data. We also display the 95\% C.L. upper bound on $c_{ee}$ by a yellow line. One can see from this figure that values of $f_{\pi}\gtrsim 1\,$TeV give already a reasonable agreement with the data, whereas for $f_{\pi}\gtrsim 1.2\,$TeV the EWPD impose no significant constraint. Therefore, in order to provide a conservative assessment of the flavor predictions of the model, we consider $f_{\pi}=1.2\,$TeV in the following.

Concerning flavor, the most relevant operators will be $\mathcal{O}_{qe}^{32\ell \ell}=\left(\bar{q}_{L}^2 \gamma_{\mu} q_L^3\right)\left(\bar{\ell}_R \gamma^{\mu} \ell_R\right)$ and $\mathcal{O}^{B_s}_1=\left(\bar{q}_L^2\gamma_{\mu} q_L^3\right)\left(\bar{q}_L^2\gamma^{\mu} q_L^3\right)$. The first one will provide the leading contribution to $R_K$ and we expect the latter to appear unavoidable if we generate the other one. Instead of performing a complete flavor analysis of the quark sector, we prefer to focus on the possible correlations between $c_1^{B_s}$ -- and thus $B_s-\bar{B}_s$ mixing -- and $R_K$. 
On the other hand, note already that a large class of potentially dangerous constraints, coming from limits on LFU violation in charged current interactions,
mediating e.g. $K$, $\pi$, and $\mu$ decays \cite{Antonelli:2010yf,Greljo:2015mma}, is fulfilled in the model at hand by construction.
In fact, the left handed charged current $\bar \ell_L \gamma_\mu \nu^\ell_L$ is mostly elementary and the light neutrino mass eigenstates contain only a negligible amount of right handed fields.
Thus, charged currents respect LFU to good approximation in the model at hand.

We evaluate $R_K$ by computing the Wilson coefficients of the   $\mathcal{O}_{9(10)}^{\ell}=\left[\bar{s}\gamma_{\alpha}P_Lb\right]\left[\bar{\ell}\gamma^{\alpha}(\gamma_5)\ell\right]$, and $\mathcal{O}_{9(10)}^{\prime \ell}=\mathcal{O}_{9(10)}^{\ell}[P_L\to P_R]$ operators from the usual $|\Delta B|=|\Delta S|=1$ Hamiltonian \cite{Ghosh:2014awa}. Note that, even though we are also generating contributions to $\mathcal{O}_{10}^{\ell}$ and $\mathcal{O}_{10}^{\prime \ell}$ that could in principle lead to large deviations with respect to the SM predictions in $B_s\to \ell^{+}\ell^{-}$ decays \cite{Bobeth:2013uxa}, 
 we  expect the largest effect to arise in the poorly measured $B_s\to e^{+}e^-$ decay, rather than in  $B_s\to \mu^{+}\mu^{-}$ \cite{CMS:2014xfa}. 

\begin{figure}[!t]
	\begin{center}
			\includegraphics[width=0.42\textwidth]{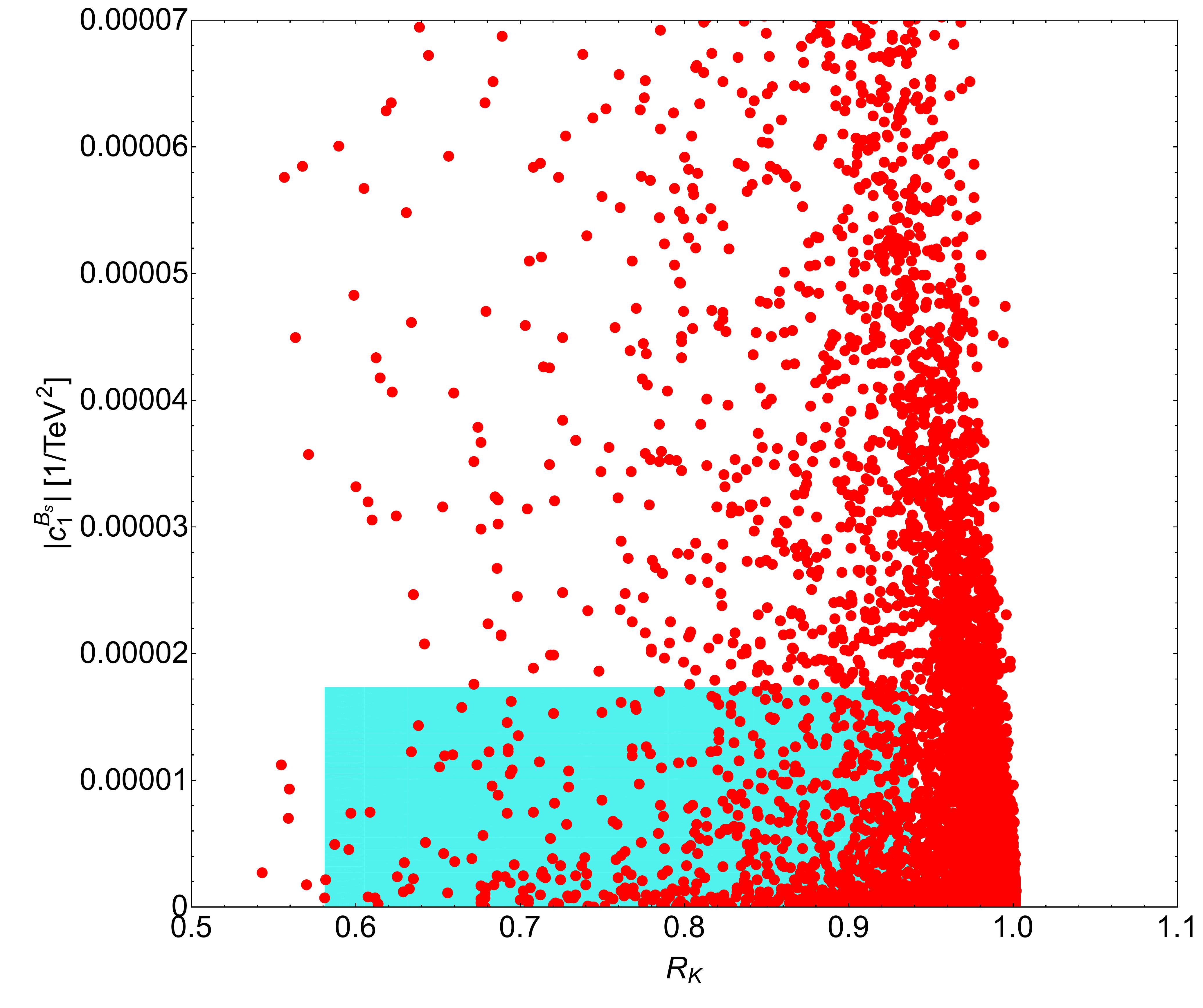}
			\caption{Value of $|c_{1}^{B_s}(m_{\rho})|$ versus $R_K$ for points reproducing the Higgs mass and within 2$\sigma$ from $B_s\to \mu^{+}\mu^-$, for $f_{\pi}=1.2\,$TeV. The blue box marks the allowed values of $R_K$ and $|c_{1}^{B_s}|$ at $95\%$ C.L..}
			\label{fig:RK}
		\end{center}
\end{figure}

We show in Figure~\ref{fig:RK} the values of $|c_{1}^{B_s}(m_{\rho})|$ versus $R_K$ for the points of the scan with the correct Higgs mass and a 2$\sigma$ agreement with the measured value of $B_s\to \mu^{+}\mu^{-}$ for  $f_{\pi}=1.2\,$TeV. The blue box represents the allowed values in the $R_K-|c_1^{B_s}|$ plane at 95\% C.L., taking into account the latest bound on $|c_1^{B_s}|\le (240\, \TeV)^{-2}$~\cite{Bevan:2014cya}.
It is clear from the plot that, even in the conservative case of $f_{\pi}=1.2\,$TeV, which guarantees the agreement with EWPD,  we can  explain the observed value of $R_K$ while not violating the bounds from  $B_s - \bar B_s$ mixing or $B_s\to \mu^{+}\mu^{-}$ for a sizable region of the parameter space. Note that an eventual experimental agreement with the SM prediction in $R_K$ would  not severely constrain the model, since values of $R_K\lesssim 1$ are also within the range of predictions. On the other hand, a measurement of $R_K>1$ could potentially rule out the model.

\section{Conclusions}
We introduced a new model of lepton-flavor in the framework of CHMs which, linked to its minimality 
has several striking features. First, it predicts a non-negligible contribution of the leptonic sector to the Higgs
potential, avoiding the necessity of ultra-light top partners. As a corollary, it even allows for more minimal 
realizations of the quark sector, compared to the models known before. Moreover, the minimality allows for a very powerful flavor protection, avoiding potentially dangerous FCNCs in the lepton sector.
Finally, although the model was not designed for it, it predicts $R_K<1$, in good agreement with the tendency
seen at LHCb, while constraints from $B$-physics as well as EWPD are met.

\begin{acknowledgments}
We are grateful to Gino Isidori for useful discussions. The research of A.C. has been supported by a Marie Sk\l{}odowska-Curie Individual Fellowship of the European Community's Horizon 2020 Framework Programme for Research and Innovation under contract number 659239 (NP4theLHC14). The research of F.G. is supported by a Marie Curie Intra European Fellowship within the EU FP7 (grant no. PIEF-GA-2013-628224). 
\end{acknowledgments}
	\bibliography{myrefs}{}

\begin{thebibliography}{30}
\expandafter\ifx\csname natexlab\endcsname\relax\def\natexlab#1{#1}\fi
\expandafter\ifx\csname bibnamefont\endcsname\relax
  \def\bibnamefont#1{#1}\fi
\expandafter\ifx\csname bibfnamefont\endcsname\relax
  \def\bibfnamefont#1{#1}\fi
\expandafter\ifx\csname citenamefont\endcsname\relax
  \def\citenamefont#1{#1}\fi
\expandafter\ifx\csname url\endcsname\relax
  \def\url#1{\texttt{#1}}\fi
\expandafter\ifx\csname urlprefix\endcsname\relax\def\urlprefix{URL }\fi
\providecommand{\bibinfo}[2]{#2}
\providecommand{\eprint}[2][]{\url{#2}}

\bibitem[{\citenamefont{Kaplan and Georgi}(1984)}]{Kaplan:1983fs}
\bibinfo{author}{\bibfnamefont{D.~B.} \bibnamefont{Kaplan}} \bibnamefont{and}
  \bibinfo{author}{\bibfnamefont{H.}~\bibnamefont{Georgi}},
  \bibinfo{journal}{Phys. Lett.} \textbf{\bibinfo{volume}{B136}},
  \bibinfo{pages}{183} (\bibinfo{year}{1984}).

\bibitem[{\citenamefont{Georgi and Kaplan}(1984)}]{Georgi:1984af}
\bibinfo{author}{\bibfnamefont{H.}~\bibnamefont{Georgi}} \bibnamefont{and}
  \bibinfo{author}{\bibfnamefont{D.~B.} \bibnamefont{Kaplan}},
  \bibinfo{journal}{Phys. Lett.} \textbf{\bibinfo{volume}{B145}},
  \bibinfo{pages}{216} (\bibinfo{year}{1984}).

\bibitem[{\citenamefont{Dimopoulos and Preskill}(1982)}]{Dimopoulos:1981xc}
\bibinfo{author}{\bibfnamefont{S.}~\bibnamefont{Dimopoulos}} \bibnamefont{and}
  \bibinfo{author}{\bibfnamefont{J.}~\bibnamefont{Preskill}},
  \bibinfo{journal}{Nucl. Phys.} \textbf{\bibinfo{volume}{B199}},
  \bibinfo{pages}{206} (\bibinfo{year}{1982}).

\bibitem[{\citenamefont{Contino et~al.}(2003)\citenamefont{Contino, Nomura, and
  Pomarol}}]{Contino:2003ve}
\bibinfo{author}{\bibfnamefont{R.}~\bibnamefont{Contino}},
  \bibinfo{author}{\bibfnamefont{Y.}~\bibnamefont{Nomura}}, \bibnamefont{and}
  \bibinfo{author}{\bibfnamefont{A.}~\bibnamefont{Pomarol}},
  \bibinfo{journal}{Nucl.Phys.} \textbf{\bibinfo{volume}{B671}},
  \bibinfo{pages}{148} (\bibinfo{year}{2003}), \eprint{hep-ph/0306259}.

\bibitem[{\citenamefont{Agashe et~al.}(2005)\citenamefont{Agashe, Contino, and
  Pomarol}}]{Agashe:2004rs}
\bibinfo{author}{\bibfnamefont{K.}~\bibnamefont{Agashe}},
  \bibinfo{author}{\bibfnamefont{R.}~\bibnamefont{Contino}}, \bibnamefont{and}
  \bibinfo{author}{\bibfnamefont{A.}~\bibnamefont{Pomarol}},
  \bibinfo{journal}{Nucl.Phys.} \textbf{\bibinfo{volume}{B719}},
  \bibinfo{pages}{165} (\bibinfo{year}{2005}), \eprint{hep-ph/0412089}.

\bibitem[{\citenamefont{Aaij et~al.}(2014)}]{Aaij:2014ora}
\bibinfo{author}{\bibfnamefont{R.}~\bibnamefont{Aaij}} \bibnamefont{et~al.}
  (\bibinfo{collaboration}{LHCb}), \bibinfo{journal}{Phys. Rev. Lett.}
  \textbf{\bibinfo{volume}{113}}, \bibinfo{pages}{151601}
  (\bibinfo{year}{2014}), \eprint{1406.6482}.

\bibitem[{\citenamefont{Hiller and Kruger}(2004)}]{Hiller:2003js}
\bibinfo{author}{\bibfnamefont{G.}~\bibnamefont{Hiller}} \bibnamefont{and}
  \bibinfo{author}{\bibfnamefont{F.}~\bibnamefont{Kruger}},
  \bibinfo{journal}{Phys. Rev.} \textbf{\bibinfo{volume}{D69}},
  \bibinfo{pages}{074020} (\bibinfo{year}{2004}), \eprint{hep-ph/0310219}.

\bibitem[{\citenamefont{Carmona and Goertz}(2015)}]{Carmona:2014iwa}
\bibinfo{author}{\bibfnamefont{A.}~\bibnamefont{Carmona}} \bibnamefont{and}
  \bibinfo{author}{\bibfnamefont{F.}~\bibnamefont{Goertz}},
  \bibinfo{journal}{JHEP} \textbf{\bibinfo{volume}{1505}}, \bibinfo{pages}{002}
  (\bibinfo{year}{2015}), \eprint{1410.8555}.

\bibitem[{\citenamefont{Maldacena}(1999)}]{Maldacena:1997re}
\bibinfo{author}{\bibfnamefont{J.~M.} \bibnamefont{Maldacena}},
  \bibinfo{journal}{Int. J. Theor. Phys.} \textbf{\bibinfo{volume}{38}},
  \bibinfo{pages}{1113} (\bibinfo{year}{1999}), \bibinfo{note}{[Adv. Theor.
  Math. Phys.2,231(1998)]}, \eprint{hep-th/9711200}.

\bibitem[{\citenamefont{Gubser et~al.}(1998)\citenamefont{Gubser, Klebanov, and
  Polyakov}}]{Gubser:1998bc}
\bibinfo{author}{\bibfnamefont{S.~S.} \bibnamefont{Gubser}},
  \bibinfo{author}{\bibfnamefont{I.~R.} \bibnamefont{Klebanov}},
  \bibnamefont{and} \bibinfo{author}{\bibfnamefont{A.~M.}
  \bibnamefont{Polyakov}}, \bibinfo{journal}{Phys. Lett.}
  \textbf{\bibinfo{volume}{B428}}, \bibinfo{pages}{105} (\bibinfo{year}{1998}),
  \eprint{hep-th/9802109}.

\bibitem[{\citenamefont{Witten}(1998)}]{Witten:1998qj}
\bibinfo{author}{\bibfnamefont{E.}~\bibnamefont{Witten}},
  \bibinfo{journal}{Adv. Theor. Math. Phys.} \textbf{\bibinfo{volume}{2}},
  \bibinfo{pages}{253} (\bibinfo{year}{1998}), \eprint{hep-th/9802150}.

\bibitem[{\citenamefont{Arkani-Hamed et~al.}(2001)\citenamefont{Arkani-Hamed,
  Porrati, and Randall}}]{ArkaniHamed:2000ds}
\bibinfo{author}{\bibfnamefont{N.}~\bibnamefont{Arkani-Hamed}},
  \bibinfo{author}{\bibfnamefont{M.}~\bibnamefont{Porrati}}, \bibnamefont{and}
  \bibinfo{author}{\bibfnamefont{L.}~\bibnamefont{Randall}},
  \bibinfo{journal}{JHEP} \textbf{\bibinfo{volume}{08}}, \bibinfo{pages}{017}
  (\bibinfo{year}{2001}), \eprint{hep-th/0012148}.

\bibitem[{\citenamefont{Fitzpatrick et~al.}(2008)\citenamefont{Fitzpatrick,
  Perez, and Randall}}]{Fitzpatrick:2007sa}
\bibinfo{author}{\bibfnamefont{A.~L.} \bibnamefont{Fitzpatrick}},
  \bibinfo{author}{\bibfnamefont{G.}~\bibnamefont{Perez}}, \bibnamefont{and}
  \bibinfo{author}{\bibfnamefont{L.}~\bibnamefont{Randall}},
  \bibinfo{journal}{Phys.Rev.Lett.} \textbf{\bibinfo{volume}{100}},
  \bibinfo{pages}{171604} (\bibinfo{year}{2008}), \eprint{0710.1869}.

\bibitem[{\citenamefont{Perez and Randall}(2009)}]{Perez:2008ee}
\bibinfo{author}{\bibfnamefont{G.}~\bibnamefont{Perez}} \bibnamefont{and}
  \bibinfo{author}{\bibfnamefont{L.}~\bibnamefont{Randall}},
  \bibinfo{journal}{JHEP} \textbf{\bibinfo{volume}{0901}}, \bibinfo{pages}{077}
  (\bibinfo{year}{2009}), \eprint{0805.4652}.

\bibitem[{\citenamefont{Csaki et~al.}(2008)\citenamefont{Csaki, Falkowski, and
  Weiler}}]{Csaki:2008zd}
\bibinfo{author}{\bibfnamefont{C.}~\bibnamefont{Csaki}},
  \bibinfo{author}{\bibfnamefont{A.}~\bibnamefont{Falkowski}},
  \bibnamefont{and} \bibinfo{author}{\bibfnamefont{A.}~\bibnamefont{Weiler}},
  \bibinfo{journal}{JHEP} \textbf{\bibinfo{volume}{0809}}, \bibinfo{pages}{008}
  (\bibinfo{year}{2008}), \eprint{0804.1954}.

\bibitem[{\citenamefont{Barbieri and Giudice}(1988)}]{Barbieri:1987fn}
\bibinfo{author}{\bibfnamefont{R.}~\bibnamefont{Barbieri}} \bibnamefont{and}
  \bibinfo{author}{\bibfnamefont{G.}~\bibnamefont{Giudice}},
  \bibinfo{journal}{Nucl.Phys.} \textbf{\bibinfo{volume}{B306}},
  \bibinfo{pages}{63} (\bibinfo{year}{1988}).

\bibitem[{\citenamefont{Gripaios et~al.}(2015)\citenamefont{Gripaios,
  Nardecchia, and Renner}}]{Gripaios:2014tna}
\bibinfo{author}{\bibfnamefont{B.}~\bibnamefont{Gripaios}},
  \bibinfo{author}{\bibfnamefont{M.}~\bibnamefont{Nardecchia}},
  \bibnamefont{and} \bibinfo{author}{\bibfnamefont{S.~A.}
  \bibnamefont{Renner}}, \bibinfo{journal}{JHEP} \textbf{\bibinfo{volume}{05}},
  \bibinfo{pages}{006} (\bibinfo{year}{2015}), \eprint{1412.1791}.

\bibitem[{\citenamefont{Niehoff et~al.}(2015)\citenamefont{Niehoff, Stangl, and
  Straub}}]{Niehoff:2015bfa}
\bibinfo{author}{\bibfnamefont{C.}~\bibnamefont{Niehoff}},
  \bibinfo{author}{\bibfnamefont{P.}~\bibnamefont{Stangl}}, \bibnamefont{and}
  \bibinfo{author}{\bibfnamefont{D.~M.} \bibnamefont{Straub}},
  \bibinfo{journal}{Phys. Lett.} \textbf{\bibinfo{volume}{B747}},
  \bibinfo{pages}{182} (\bibinfo{year}{2015}), \eprint{1503.03865}.

\bibitem[{\citenamefont{Crivellin
  et~al.}(2015{\natexlab{a}})\citenamefont{Crivellin, D'Ambrosio, and
  Heeck}}]{Crivellin:2015mga}
\bibinfo{author}{\bibfnamefont{A.}~\bibnamefont{Crivellin}},
  \bibinfo{author}{\bibfnamefont{G.}~\bibnamefont{D'Ambrosio}},
  \bibnamefont{and} \bibinfo{author}{\bibfnamefont{J.}~\bibnamefont{Heeck}},
  \bibinfo{journal}{Phys.Rev.Lett.} \textbf{\bibinfo{volume}{114}},
  \bibinfo{pages}{151801} (\bibinfo{year}{2015}{\natexlab{a}}),
  \eprint{1501.00993}.

\bibitem[{\citenamefont{Crivellin
  et~al.}(2015{\natexlab{b}})\citenamefont{Crivellin, D'Ambrosio, and
  Heeck}}]{Crivellin:2015lwa}
\bibinfo{author}{\bibfnamefont{A.}~\bibnamefont{Crivellin}},
  \bibinfo{author}{\bibfnamefont{G.}~\bibnamefont{D'Ambrosio}},
  \bibnamefont{and} \bibinfo{author}{\bibfnamefont{J.}~\bibnamefont{Heeck}},
  \bibinfo{journal}{Phys. Rev.} \textbf{\bibinfo{volume}{D91}},
  \bibinfo{pages}{075006} (\bibinfo{year}{2015}{\natexlab{b}}),
  \eprint{1503.03477}.

\bibitem[{\citenamefont{Sierra et~al.}(2015)\citenamefont{Sierra, Staub, and
  Vicente}}]{Sierra:2015fma}
\bibinfo{author}{\bibfnamefont{D.~A.} \bibnamefont{Sierra}},
  \bibinfo{author}{\bibfnamefont{F.}~\bibnamefont{Staub}}, \bibnamefont{and}
  \bibinfo{author}{\bibfnamefont{A.}~\bibnamefont{Vicente}},
  \bibinfo{journal}{Phys. Rev.} \textbf{\bibinfo{volume}{D92}},
  \bibinfo{pages}{015001} (\bibinfo{year}{2015}), \eprint{1503.06077}.

\bibitem[{\citenamefont{Celis et~al.}(2015)\citenamefont{Celis, Fuentes-Martin,
  Jung, and Serodio}}]{Celis:2015ara}
\bibinfo{author}{\bibfnamefont{A.}~\bibnamefont{Celis}},
  \bibinfo{author}{\bibfnamefont{J.}~\bibnamefont{Fuentes-Martin}},
  \bibinfo{author}{\bibfnamefont{M.}~\bibnamefont{Jung}}, \bibnamefont{and}
  \bibinfo{author}{\bibfnamefont{H.}~\bibnamefont{Serodio}},
  \bibinfo{journal}{Phys. Rev.} \textbf{\bibinfo{volume}{D92}},
  \bibinfo{pages}{015007} (\bibinfo{year}{2015}), \eprint{1505.03079}.

\bibitem[{\citenamefont{Falkowski et~al.}(2015)\citenamefont{Falkowski,
  Nardecchia, and Ziegler}}]{Falkowski:2015zwa}
\bibinfo{author}{\bibfnamefont{A.}~\bibnamefont{Falkowski}},
  \bibinfo{author}{\bibfnamefont{M.}~\bibnamefont{Nardecchia}},
  \bibnamefont{and} \bibinfo{author}{\bibfnamefont{R.}~\bibnamefont{Ziegler}}
  (\bibinfo{year}{2015}), \eprint{1509.01249}.

\bibitem[{\citenamefont{Raidal et~al.}(2008)}]{Raidal:2008jk}
\bibinfo{author}{\bibfnamefont{M.}~\bibnamefont{Raidal}} \bibnamefont{et~al.},
  \bibinfo{journal}{Eur. Phys. J.} \textbf{\bibinfo{volume}{C57}},
  \bibinfo{pages}{13} (\bibinfo{year}{2008}), \eprint{0801.1826}.

\bibitem[{\citenamefont{Antonelli et~al.}(2010)}]{Antonelli:2010yf}
\bibinfo{author}{\bibfnamefont{M.}~\bibnamefont{Antonelli}}
  \bibnamefont{et~al.} (\bibinfo{collaboration}{FlaviaNet Working Group on Kaon
  Decays}), \bibinfo{journal}{Eur. Phys. J.} \textbf{\bibinfo{volume}{C69}},
  \bibinfo{pages}{399} (\bibinfo{year}{2010}), \eprint{1005.2323}.

\bibitem[{\citenamefont{Greljo et~al.}(2015)\citenamefont{Greljo, Isidori, and
  Marzocca}}]{Greljo:2015mma}
\bibinfo{author}{\bibfnamefont{A.}~\bibnamefont{Greljo}},
  \bibinfo{author}{\bibfnamefont{G.}~\bibnamefont{Isidori}}, \bibnamefont{and}
  \bibinfo{author}{\bibfnamefont{D.}~\bibnamefont{Marzocca}},
  \bibinfo{journal}{JHEP} \textbf{\bibinfo{volume}{07}}, \bibinfo{pages}{142}
  (\bibinfo{year}{2015}), \eprint{1506.01705}.

\bibitem[{\citenamefont{Ghosh et~al.}(2014)\citenamefont{Ghosh, Nardecchia, and
  Renner}}]{Ghosh:2014awa}
\bibinfo{author}{\bibfnamefont{D.}~\bibnamefont{Ghosh}},
  \bibinfo{author}{\bibfnamefont{M.}~\bibnamefont{Nardecchia}},
  \bibnamefont{and} \bibinfo{author}{\bibfnamefont{S.}~\bibnamefont{Renner}},
  \bibinfo{journal}{JHEP} \textbf{\bibinfo{volume}{1412}}, \bibinfo{pages}{131}
  (\bibinfo{year}{2014}), \eprint{1408.4097}.

\bibitem[{\citenamefont{Bobeth et~al.}(2014)\citenamefont{Bobeth, Gorbahn,
  Hermann, Misiak, Stamou, and Steinhauser}}]{Bobeth:2013uxa}
\bibinfo{author}{\bibfnamefont{C.}~\bibnamefont{Bobeth}},
  \bibinfo{author}{\bibfnamefont{M.}~\bibnamefont{Gorbahn}},
  \bibinfo{author}{\bibfnamefont{T.}~\bibnamefont{Hermann}},
  \bibinfo{author}{\bibfnamefont{M.}~\bibnamefont{Misiak}},
  \bibinfo{author}{\bibfnamefont{E.}~\bibnamefont{Stamou}}, \bibnamefont{and}
  \bibinfo{author}{\bibfnamefont{M.}~\bibnamefont{Steinhauser}},
  \bibinfo{journal}{Phys. Rev. Lett.} \textbf{\bibinfo{volume}{112}},
  \bibinfo{pages}{101801} (\bibinfo{year}{2014}), \eprint{1311.0903}.

\bibitem[{\citenamefont{Khachatryan et~al.}(2015)}]{CMS:2014xfa}
\bibinfo{author}{\bibfnamefont{V.}~\bibnamefont{Khachatryan}}
  \bibnamefont{et~al.} (\bibinfo{collaboration}{LHCb, CMS}),
  \bibinfo{journal}{Nature} \textbf{\bibinfo{volume}{522}}, \bibinfo{pages}{68}
  (\bibinfo{year}{2015}), \eprint{1411.4413}.

\bibitem[{\citenamefont{Bevan et~al.}(2014)}]{Bevan:2014cya}
\bibinfo{author}{\bibfnamefont{A.}~\bibnamefont{Bevan}} \bibnamefont{et~al.}
  (\bibinfo{year}{2014}), \eprint{1411.7233}.

\end{thebibliography}
\end{document}